\documentclass[aps,prl,twocolumn,groupedaddress]{revtex4-1}
\usepackage[latin9]{inputenc}
\usepackage{verbatim}
\usepackage{graphicx}
\usepackage{xargs}[2008/03/08]
\usepackage{times}
\usepackage{amsmath}

\begin{document}
	
	\preprint{}
	
	\title{Measuring the single-photon temporal-spectral wave function}
	
	
	\author{Alex O. C. Davis$^{1}$}
	\author{Val\'{e}rian Thiel$^{1}$}
	\author{Micha\l{} Karpi\'{n}ski{$^{1,2}$}}
	\author{Brian J. Smith{$^{1,3}$}}
	\affiliation{{$^{1}$}Clarendon Laboratory, University of Oxford, Parks Road, Oxford, OX1 3PU, UK}
	\affiliation{{$^{2}$}Faculty of Physics, University of Warsaw, Pasteura 5, 02-093 Warszawa, Poland}
	\affiliation{{$^{3}$}Department of Physics and Oregon Center for Optical, Molecular, and Quantum Science, University of Oregon, Eugene, Oregon 97403, USA }
	
	
	\date{\today}
	
	\begin{abstract}
		Temporal-spectral modes of light provide a fundamental window into the nature of quantum systems and offer robust means for information encoding. Methods to precisely characterize the temporal-spectral state of light at the single-photon level thus play a central role in understanding single-photon sources and their applications in emerging optical quantum technologies. Here we demonstrate an optical reference-free method, which melds techniques from ultrafast metrology and single-photon spectral detection, to characterize the temporal-spectral state of single photons.
	\end{abstract}

	
	\maketitle
	
	
	
	
	The quantum state of a physical system, expressed as the wave function in the case of pure states, provides a complete description of the system and allows statistical predictions of measurement outcomes performed on it. Characterization of the quantum state plays a central role in the foundations of quantum physics and applications in quantum information science \cite{lvovsky:09,o2009photonic}. For a single photon, the quantum state is given by the electromagnetic field mode it occupies, which can be viewed as the photon wave function \cite{birula:94,sipe:95,smith:07}. The various independent degrees of freedom that comprise the modes of light can be used to encode information in the electromagnetic field, namely transverse position-momentum \cite{sasada:03}, time-frequency \cite{thiel:17}, and polarization \cite{pittman:02}. Many preliminary demonstrations of quantum optical technologies have utilized polarization, path or transverse-spatial mode encoding. These degrees of freedom are limited to relatively few quantum bits (qubits) that can be practically addressed within an integrated-optics platform, where high-stability, low-loss multi-photon interference, necessary for many optical quantum technologies, can occur.  Recent research within quantum optics and quantum information science has focused on time-frequency encoding in ultrashort pulsed modes, stemming, in part, from the compatibility of pulsed modes with integrated optical platforms and the large information capacity of the time-frequency degree of freedom \cite{nunn:13, brecht:15}. Ultrashort optical pulses are a key resource in modern physics and technology with applications ranging from precision metrology \cite{hentschel:01} and spectroscopy \cite{cundiff:13} to communications and control \cite{weiner:11}. Their advantages are now being recognized for applications within optical quantum technologies, including quantum information processing \cite{roslund:14, brecht:15}, quantum-enhanced sensing \cite{giovannetti:11} and quantum cryptography \cite{dixon:10}. 
	
	
	For a photon occupying given transverse-spatial and polarization modes, characterization of the temporal-spectral electromagnetic field mode completely determines the single-photon state. There are several well-established techniques that enable measurement of the pulse-mode structure of high-intensity optical fields \cite{walmsley:09}. Among these are methods that utilize optical nonlinear interactions in the test field such as frequency-resolved optical gating (FROG) \cite{kane:93} and spectral phase interferometry for direct electric field reconstruction (SPIDER) \cite{iaconis:98}. However, at the single-photon level these techniques are challenging to realize due to the weakness of optical nonlinear effects. Methods to measure the single-photon time-frequency (TF) state based on interference with known reference pulses in either a linear optical framework \cite{wasilewski:07, polycarpou:12, qin:15}, or a nonlinear optical regime \cite{ansari:16}, have been recently demonstrated. These approaches require stable, tunable, well-characterized reference pulses for reliable measurements. Furthermore, techniques that utilize nonlinear interaction between a reference field and the single-photon state under examination require group-velocity- and phase-matching conditions to be satisfied, impeding the ability to characterize pulses over a broad spectral range in a single experimental configuration. 
	
	In this Letter we present a self-referencing linear-optical method to completely characterize the temporal-mode state of broadband single photons. This approach, based on spectral shearing interferometry \cite{iaconis:98}, enables characterization of the single-photon spectral-temporal wave function using electro-optic spectral shearing \cite{wright:17} and single-photon spectral measurements \cite{davis:17}. The fact that no additional optical fields besides the test pulse are utilized implies that this technique is applicable across a broad range of central wavelengths. Furthermore, this method does not require scanning of the reference field to achieve complete state reconstruction, which enables real-time feedback for source optimization.
	

	The state of an ensemble of identically prepared single photons is given by the electromagnetic field mode function that the photons occupy. Here we presume that the transverse-spatial and polarization state of the photons are known and focus on the pulse mode structure, encapsulated by the complex-valued temporal-mode function, $\psi(t)$ for a pure state. The temporal mode function can be viewed as the temporal wave function, which is a solution to the wave equation in the slowly-varying-envelope approximation \cite{diels:06, brecht:15}. In the following we utilize the spectral representation of the temporal pulse mode amplitude $\psi(t)$, given by its Fourier transform, $\tilde{\psi}(\omega)=\mathcal{FT}\{\psi(t)\}$, since the techniques discussed involve measurements in the spectral domain. Thus the single-photon component of the electromagnetic field when there is a single photon occupying the spectral-temporal mode, $\tilde{\psi}(\omega)$, can be expressed by
	\begin{equation}
	| 1_{\psi} \rangle =  \int \textrm{d}\omega~ \tilde{\psi}(\omega) {\hat{a}}^{\dagger}(\omega) | 0 \rangle,
	\label{1photonstate}
	\end{equation}
	\noindent where ${\hat{a}}^{\dagger}(\omega)$ is a field operator creating a single monochromatic photon of frequency $\omega$ and $| 0 \rangle$ is the vacuum state of the field. 
	
	%

	Complete characterization of the single-photon state in this case amounts to measurement of the complex-valued spectral amplitude $\tilde{\psi}(\omega)$, which requires the reconstruction of both the magnitude of the spectral amplitude $|\tilde{\psi}(\omega)|$ and the spectral phase $\phi(\omega) = \mathrm{Arg}[\tilde{\psi}(\omega)]$. Spectral amplitude can be determined directly through acquisition of the optical spectrum $S(\omega) = |\tilde{\psi}(\omega)|^2$, as recently demonstrated by frequency-to-time mapping and time-resolved detection in both the visible and telecommunications spectral ranges \cite{avenhaus:09, davis:17}. Determination of the spectral phase $\phi(\omega)$ is more challenging and at optical frequencies a non-stationary process is necessary \cite{wong:94}. Methods to extract it often rely on the principle of spectral interferometry \cite{walmsley:09}, where the pulse under test is interfered with either a reference pulse, or in the case of self-referencing methods, a modified copy of itself. This is followed by spectrally-resolved detection, allowing reconstruction of the spectral phase with Fourier analysis. 
	
	
	\begin{figure}
		\includegraphics[width=.9\linewidth]{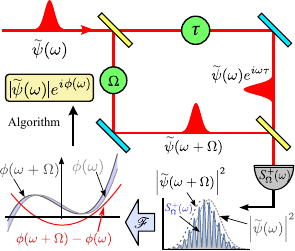}
		\caption{\textbf{Measurement principle.} \emph{Clockwise:} The test pulse, in a spectral mode determined by the unknown function $\tilde{\psi}(\omega)$, enters a Mach-Zehnder interferometer in which one arm receives a spectral shear $\Omega$ via an EOM and the other a relative time delay $\tau$. Spectrally resolved detection reveals an interference pattern $S_\Omega^+$. Fourier analysis of the observed interferences enables the extraction of $\phi(\omega+\Omega)-\phi(\omega)$ (solid red curve) which enables determination of $\phi(\omega)$. Together with the spectral amplitude measurements, the full amplitude and phase of the single photon-occupied mode can then be computationally reconstructed.}\label{fig:schem}
	\end{figure}
	
	\begin{figure}
		\includegraphics[width=.85\linewidth]{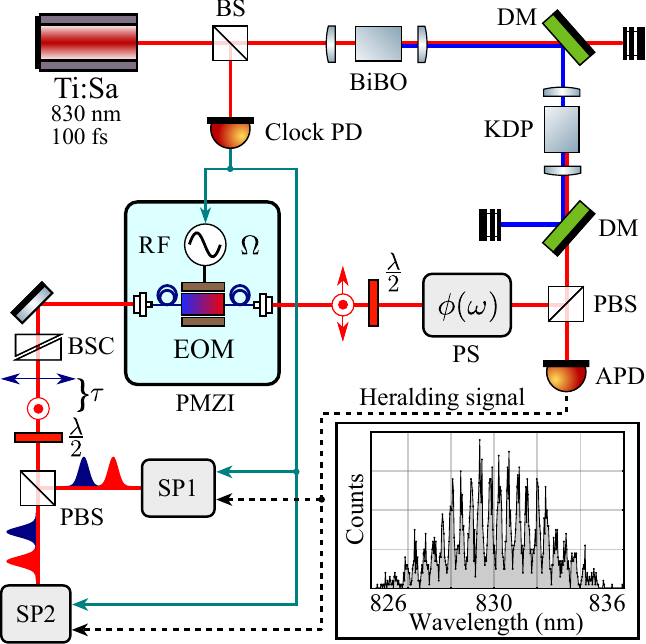}
		\caption{\textbf{Experimental set up}. Second harmonic of a Ti:Sapph laser beam (Spectra-Physics Tsunami, 830~nm center wavelength, 80~MHz repetition rate) pumps a spontaneous parametric down-conversion source of broadband heralded single photons. BS, low reflectivity pick-off beam splitter. Clock PD, fast photodiode used to trigger experiment clock. DM, dichroic mirror. PBS,  polarizing beam splitter. APD, avalanche photodiode used to herald the existence of a single photon in the signal arm. PS, pulse shaper. PMZI, polarization Mach-Zehnder interferometer. $\lambda/2$, half-wave plate. EOM, electro-optic phase modulator. RF, 10~GHz radio frequency source.  BSC, Babinet-Soleil compensator. SP, spectrometer. Electro-optically modulated (unmodulated) polarization is indicated by in-plane (out-of-plane) arrows. Inset: typical single-photon spectral interferogram at the output of the EOSI showing a 20 seconds acquisition.}
		\label{fig:experiment}
	\end{figure}
	
	A common implementation of the self-referenced approach to spectral phase reconstruction is spectral shearing interferometry \cite{walmsley:09}. The protocol for obtaining the spectral phase is outlined in Fig \ref{fig:schem}. This involves splitting the pulse with a 50:50 beam splitter, applying a constant spectral translation, or spectral shear, $\Omega$, to one of the copies and a temporal delay to the other, recombining the two at a beam splitter, and measuring the resulting spectral interference pattern. For a single photon pulse in a mode given by $\tilde{\psi}(\omega)$ the intensities at the outputs of the interferometer are:
	\begin{align} 
	S_\Omega^{\pm}(\omega,\tau)= &\frac{1}{4}\left\{S(\omega)+S(\omega+\Omega) \right. \\ &\pm \left. 2\mbox{Re}\left[\tilde{\psi}(\omega)\tilde{\psi}^*(\omega+\Omega)e^{i\omega\tau}\right] \right\}, \nonumber
	\end{align}
	where $S(\omega)$ is the spectral intensity of the original pulse. The interferogram $S_\Omega^{\pm}(\omega,\Omega,\tau)$ (Fig.\ \ref{fig:experiment} inset) contains information about the spectral phase $\phi(\omega)-\phi(\omega+\Omega)$ in the interferogram. If $\Omega$ and $\tau$ are known, the spectral phase can be extracted using Fourier transform-based algorithms \cite{walmsley:09,dorrer:01}. Arbitrary spectral amplitude can in principle be completely reconstructed by monitoring both outputs of the interferometer.
	
	A method of obtaining the spectral shift $\Omega$ that is independent of the intensity of the test pulse, and therefore feasible at the quantum level, is direct linear temporal phase modulation. Recently there have been several demonstrations of spectral shifts of quantum light pulses by direct temporal phase modulation, optomechanically \cite{fan:16} and by using cross phase modulation driven by a strong pump field \cite{matsuda:16}. Frequency translation of an optical pulse train can also be achieved in this way electro-optically \cite{wright:17}, forming the basis of electro-optic spectral shearing interferometry (EOSI), which has been succesfully used to characterize classical light pulses \cite{dorrer:03}. In our implementation we use the electro-optic technique, with a lithium-niobate-waveguide electro-optic phase modulator \cite{wright:17}. Electro-optic spectral shearing offers several advantages that make it suitable for spectral shearing interferometry. The spectral shear can be realized uniformly over a large wavelength range using a single device, and is largely independent of the incident pulse shape. Hence only basic knowledge about the input mode (such as wavelength compatibility with the set-up) is required and the same experimental configuration can accurately characterize any of a diverse range of input pulses  \cite{PRA:17}.
	
	The experimental setup for the pulse mode reconstruction is depicted in Fig \ref{fig:experiment}. The heralded single photons used in this experiment were broadband ($830$~nm central wavelength with $8$~nm full width at half-maximum (FWHM) bandwidth), in a spectrally pure state and were generated by frequency-degenerate spontaneous parametric downconversion inside a potassium dihydrogen phosphate (KDP) crystal pumped at $80$~MHz repetition rate \cite{mosley:08}. These heralded photons were measured to have a $g^{(2)}$ value of $0.06\pm0.04$. The signal photon, after separation from the idler at a polarizing beam-splitter (PBS), was routed using a single-mode fiber to a home-built pulse shaper (PS) capable of imprinting arbitrary spectral phase profiles. As well as compensating for the spectral phase introduced by the device itself prior to injection into the interferometer, the PS enables the device to be tested with pulses with various phase profiles, where the phase or its derivatives may exhibit pathological behaviour \cite{PRA:17}. The heralding idler photon was detected by a single-mode-fiber-coupled single-photon counting module. After exiting the PS the signal photon was rotated 45$^\circ$ in polarization and injected into a polarization Mach-Zehnder interferometer (PMZI), where the two arms of the interferometer are the two collinear polarization modes in a birefringent set-up, closed by a polarizing beam splitter oriented at 45$^\circ$ to the polarization axes which thus acts as a 50:50 beam splitter. This common-path arrangement provides the stability required for long acquisition times, since the full reconstruction of a single-photon wave function over a high-dimensional Hilbert space requires many copies of the original pulse \cite{lvovsky:09}. Within the PMZI the beam was coupled into a polarization-maintaining fiber-coupled electro-optic phase modulator (EOM) driven by a $10$~GHz sinusoidal radio-frequency (RF) signal synchronized with the pump laser pulse train, where one of the polarization components was spectrally sheared by $0.58 \pm 0.02$~nm. Subsequently a calcite Babinet-Soleil compensator was used to adjust the time delay $\tau$ between the two orthogonally polarized components. The independently-determined value of $\tau$ is crucial for accurate spectral phase reconstruction. We used an interferometric method with bright classical light to determine $\tau$ \cite{PRA:17}.  The polarization components were interfered using a half-wave-plate and a PBS, closing the PMZI, with PBS outputs fiber-coupled and directed to two time-resolved single photon counting spectrometers (TRSPSs) for acquisition of the heralded single-photon output spectra \cite{davis:17}.
	
	
	\begin{figure}[t]
		\includegraphics[width=\linewidth]{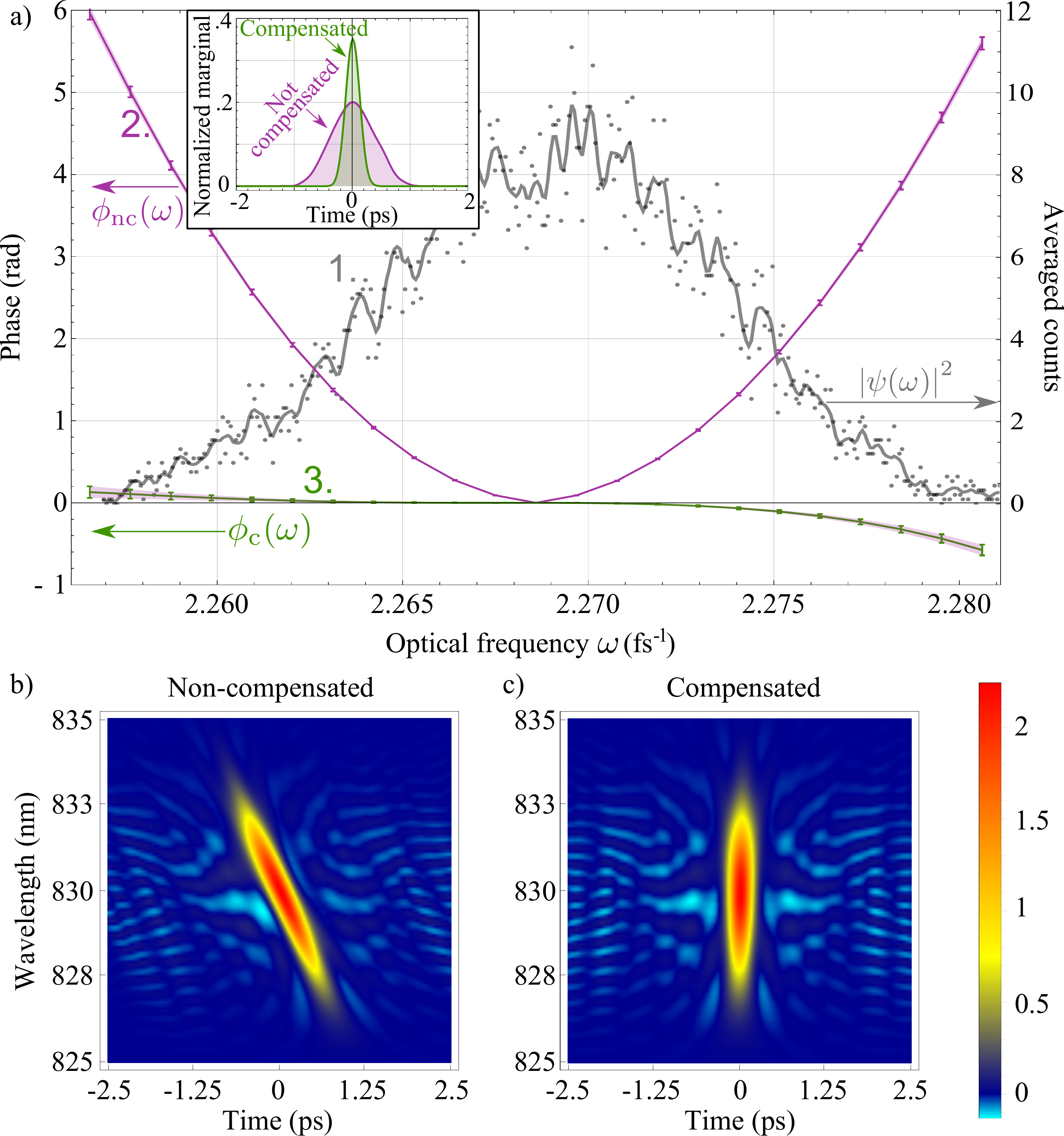}
		\caption{Top: reconstructed phases for unshaped ($\phi_\textrm{nc}$, 2.) and compensated for quadratic phase ($\phi_\textrm{c}$, 3.) cases. The greyscale plot (1.) assigned to the right-hand scale represents the single photon spectral intensity averaged over multiple runs. Inset: temporal wavefunction. Bottom: chronocyclic Wigner functions of b) unshaped pulse and c) compensated for excess quadratic phase.}
		\label{Phase}
	\end{figure}
	Figure \ref{fig:experiment} (inset) shows a raw interferogram typical of those used to perform spectral-temporal state reconstruction and incorporates approximately 20 seconds of data acquisition. Summing the interferograms at both outputs cancels the fringe pattern, allowing measurement of the single photon spectral envelope.
	
	\begin{figure}[t]
		\includegraphics[width=1\linewidth]{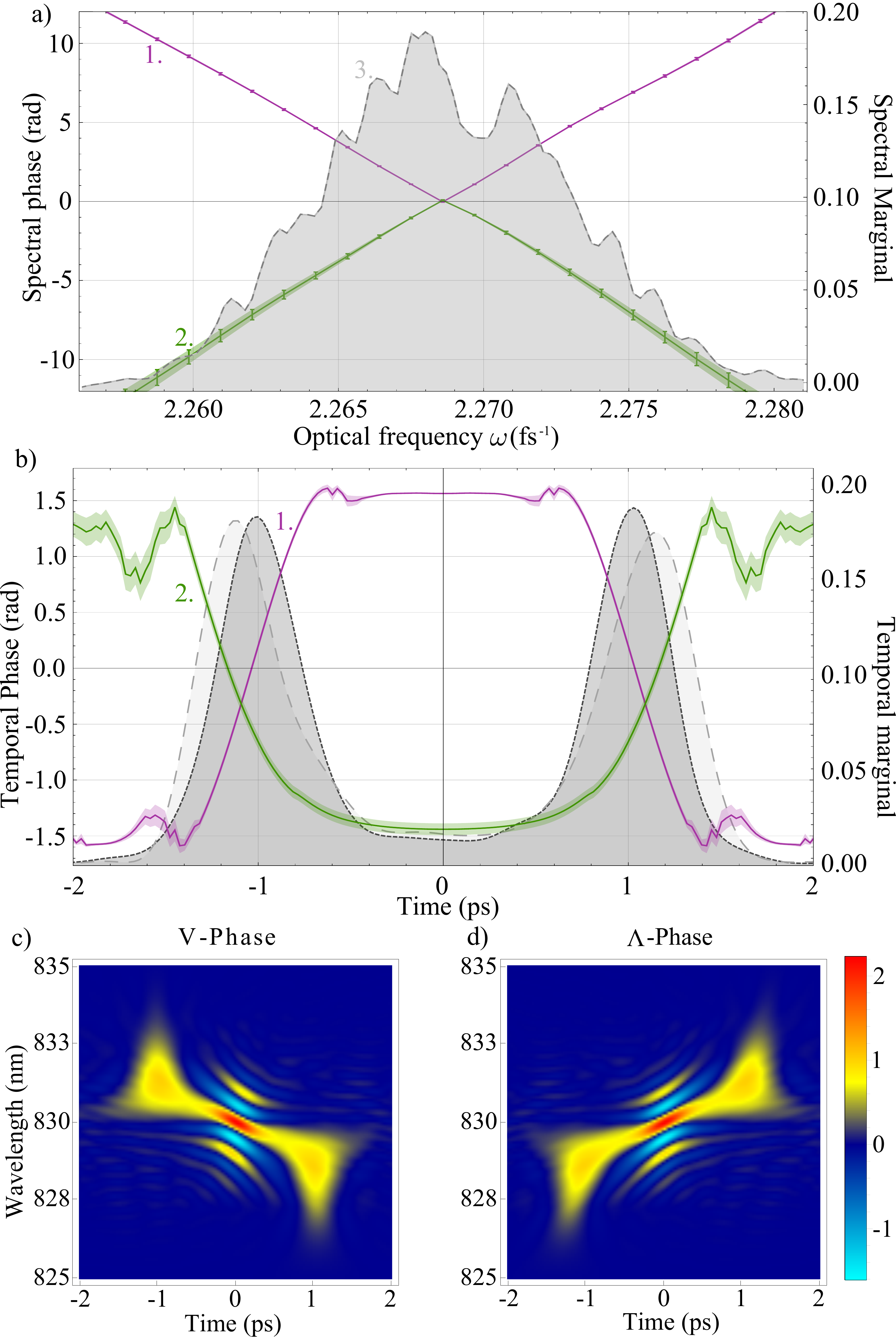}
		\caption{a) Reconstructed spectral phase profiles of pulses with applied V-phase (1. purple) and $\Lambda$-phase (2. green), with the single photon spectrum shown in greyscales (3.). b) Reconstructed temporal phases for V-phase (1. purple) and $\Lambda$-phase (2. green), as well as the temporal profile, showing binodal temporal distribution. c) Chronocyclic Wigner functions of V-phase pulse and d) $\Lambda$-phase pulse.}
		\label{fyp_lambda}
	\end{figure}
	
	To demonstrate the wide applicability of this method, we characterize the path leading up to the PMZI and compensate using the pulse shaper. Figure \ref{Phase} shows the spectral phases of two pulses with similar, approximately Gaussian spectral intensities. One of these pulses is left with its original spectral phase profile, which includes the effect of several meters of dispersive glass fiber leading into the interferometer. The device therefore reconstructs a large positive quadratic spectral phase of $(8.7\pm 0.1)\cdot 10^4$ fs$^2$ and a small cubic component of $(5.0\pm1.0)\cdot 10^5$ fs$^3$. For the second pulse train, the previously reconstructed quadratic spectral phase was imprinted on the pulse train with opposite sign using the shaper to cancel this measured phase and the experiment was repeated.
	The reconstructed phases are shown in Fig. \ref{Phase}. The blue curve represents the unshaped single photon while the green curve shows the compensation of the quadratic phase by the pulse shaper, while the uncorrected, cubic component still remains. The extracted quadratic phase coefficient in the corrected cases is just $(-1.3 \pm 0.9)\cdot 10^3$ fs$^2$, indicating close agreement between the phase applied on the pulse shaper and that measured by the phase reconstruction.
	Using the spectrum also shown in Fig. \ref{Phase}a) and the reconstructed phases, it is possible to calculate the temporal wavefunction $\psi(t)$ of the single photons (inset), which shows that the compensated single photon pulse is indeed compressed to its transform limit duration. The amount of remaining cubic phase is small enough that is does not significantly change the temporal waveform.
	Finally, Fig. \ref{Phase}b) and c) show the chronocylic Wigner distributions for these two pulses. This representation constitutes a quasiprobability distribution in the time-frequency space of the pulse and allows spectral-temporal correlations, such as those brought about by spectral phase, to be visualized \cite{paye:92}. The delay in the arrival time in the shorter-wavelength frequencies, brought about by the second-order spectral phase, can be seen clearly in this image.
	
	As a further demonstration of the versatility of the single-photon EOSI, the pulse shaper was used to manipulate single photons into one of the two states with the spectral phase profiles shown in Fig. \ref{fyp_lambda}, termed ``V-phase" and ``$\Lambda$-phase". The pulses produced by these unitary spectral phase manipulations have a binodal temporal intensity profile, caused by a relative delay between the high-frequency and low-frequency spectral components (with opposite sign for the two different phase profiles). When the gradient of the phase is much greater than the duration of the original pulse, these two peaks can be very well separated in time \cite{compressionnote}. The V-phase pulse and the $\Lambda$-phase pulse thus have the property that they have identical spectral and temporal intensities, but are nearly orthogonal, despite being created from the same initial resource through phase-only operations. Distinguishing these two orthogonal states can therefore only be achieved with phase-sensitive measurements. 
	Fig. \ref{fyp_lambda}(a) shows the reconstructed phase profiles of the two pulses, providing excellent agreement with the phase written onto the pulse shaper. The reconstructed linear phase coefficients are found to be $1050 \pm 100$ fs for the V-phase and $-1100 \pm 200$ fs for the $\Lambda$-phase. Using again the experimental single photon spectrum, the reconstructed binodal temporal intensity distributions are also shown on Fig. \ref{fyp_lambda}(b) along with the chronocyclic Wigner functions of the two pulses (c and d), showing the high-frequency and low-frequency peaks and the coherences between them. Although the marginals are indistinguishable, the modes are indeed nearly orthogonal, and computing the mode overlap between the two reconstructed modes yields a value of $0.06 \pm 0.01$.
	
	In summary, we have demonstrated a self-referencing technique to reconstruct the spectral-temporal state of a pulsed single photon. This method is applicable across a wide range of wavelengths and pulse characteristics, and no reference fields are required. The only \emph{a priori} knowledge required is that necessary to ensure technical compatibility with the optics and detectors. Full characterization of an arbitrary pure state can be performed in a single run of measurements, without reconfiguration of the apparatus. Furthermore, the spectral shear $\Omega$ can easily be reconfigured, enabling a more robust reconstruction of complicated or even partially mixed pulse trains \cite{bourassimbouchet:13}. Our method is readily generalized to multi-photon states in separate spatial or polarization modes, enabling the full reconstruction of multi-photon states featuring high-dimensional multipartite entanglement in the TF basis. Our device has wide applications in single-photon source and detector characterization and quantum metrology. Characterization of single-photon sources of typical luminosity is rapid enough to enable real-time feedback in source optimization. We anticipate that future experiments will also utilize this technology for more diverse purposes such as spectral-temporal entanglement characterization and state purity determination. 
	
	\begin{acknowledgments}
		We are grateful to C. Dorrer, C. Radzewicz, M. G. Raymer, and I. A. Walmsley for fruitful discussions. This project has received funding from the European Union's Horizon 2020 research and innovation programme under Grant Agreement No. 665148, the United Kingdom Defense Science and Technology Laboratory (DSTL) under contract No. DSTLX-100092545, and the National Science Foundation under Grant No. 1620822. This work was partially funded by the HOMING programme of the Foundation for Polish Science (project no. Homing/2016-1/4) co-financed by the European Union under the European Regional Development Fund.
	\end{acknowledgments}
	
	\section*{References}
	\bibliography{bibliographyPRL}

\begin{thebibliography}{37}
\bibitem [{com()}]{compressionnote}%
  \BibitemOpen
  \bibinfo {note} {This implies that the pulses are transform limited. The
  phase was thus applied on top of the correction of quadratic phase that
  compresses the temporal wavefunction to the maximum limit.}\BibitemShut
  {Stop}%
\end{thebibliography}

\end{document}